\documentclass[runningheads]{llncs}
\usepackage{graphicx}
\usepackage[misc,geometry]{ifsym} 
\usepackage{xcolor}
\usepackage{amsfonts}
\usepackage{caption}

\begin{document}
\title{Region Growing with Convolutional Neural Networks for Biomedical Image Segmentation }
%
\titlerunning{Region Growing with Convolutional Neural Networks}
%

\author{John H. Lagergren$^{\left(\textrm{\Letter}\right)}$ \and Erica M. Rutter \and Kevin B. Flores}
\authorrunning{J. H. Lagergren et al.}
\institute{Center for Research in Scientific Computation, Department of Mathematics,
\\ 
North Carolina State University
\\
\email{\{jhlagerg, erutter, kbflores\}@ncsu.edu}   }
\maketitle              
\begin{abstract}
In this paper we present a methodology that uses convolutional neural networks (CNNs) for segmentation by iteratively growing predicted mask regions in each coordinate direction. The CNN is used to predict class probability scores in a small neighborhood of the center pixel in a tile of an image. We use a threshold on the CNN probability scores to determine whether pixels are added to the region and the iteration continues until no new pixels are added to the region. Our method is able to achieve high segmentation accuracy and preserve biologically realistic morphological features while leveraging small amounts of training data and maintaining computational efficiency. Using retinal blood vessel images from the DRIVE database we found that our method is more accurate than a fully convolutional semantic segmentation CNN for several evaluation metrics. 

\end{abstract}
\section{Introduction}


The use of convolutional neural networks (CNNs) for image segmentation has been widely adopted for the development of automated methods in the detection and analysis of diverse biomedical phenomena \cite{ciresan2012neuronal,khan2018lung,ronneberger2015unet,rouhi2015breast,valen2016cells}. Morphological attributes of the segmentation masks (e.g., shape, size, emergent patterns, etc.) can be used for the computer-assisted diagnosis, evaluation, and treatment of various diseases. A representative example is the DRIVE (Digital Retinal Images for Vessel Extraction) database \cite{staal2004vessel}, which was designed to evaluate the accuracy of methods for retinal blood vessel segmentation. Morphological features of retinal vasculature segmentations, such as vessel width, tortuosity, and abnormal branching patterns, can aid in the screening of cardiovascular and ophthalmologic diseases like diabetic retinopathy, a leading cause of blindness \cite{staal2004vessel}. Thus, a key challenge in applying CNNs to biomedical image segmentation is to produce segmentations that are both accurate and realistic enough to preserve biological features from which experts can draw meaningful conclusions for patient care. \\
\indent One primary reason that it is challenging to use CNN segmentation predictions to draw meaningful conclusions is that they may produce patchy output probability maps (Supplementary Figure 1). The lack of contiguity in the resulting segmentation mask, which is caused by thresholding the probability map, is biologically unrealistic and therefore may not be predictive of morphological features important for diagnosis. Here, we present a novel CNN-based algorithm for biomedical image segmentation that better ensures contiguity of the predicted masks by iteratively growing and connecting regions. This work extends previous region growing algorithms \cite{te_brake_segmentation_2001,wei_mammogram_2012} by replacing pixel selection criteria with CNN output probabilities. We compare the performance of our method on retinal images from the DRIVE database with a baseline CNN for semantic segmentation (U-net \cite{ronneberger2015unet}).

\noindent \textbf{Related Work: Region Growing.} Region growing is a segmentation approach in which regions of an image are segmented by grouping together neighboring pixels that are similar to initial seed points. Adjacent pixels are compared for similarity and grouped together if their similarity exceeds some threshold. By iteratively grouping neighboring pixels, a region is grown until no similar pixels remain. Traditionally, similarity criteria are based on features such as pixel intensity, surrounding texture, color, etc. \cite{te_brake_segmentation_2001,wei_mammogram_2012}. Neural networks and region growing have been combined previously \cite{khan2018lung,rouhi2015breast}. However, in \cite{khan2018lung} neural networks were not used to make decisions about adding pixels to the region, and in \cite{rouhi2015breast} neural networks were used to choose image threshold values for traditional similarity criteria. Our method is novel in that it combines region growing with CNNs to directly predict the segmentation mask. 

\noindent \textbf{Related Work: CNN Methods.} Approaches to using CNNs for biomedical image segmentation range from fully convolutional methods \cite{long2015fully,ronneberger2015unet} which classify pixels in large tiles, to patch-based methods \cite{ciresan2012neuronal,valen2016cells} which classify pixels individually from local patches surrounding the pixel. The computational time for fully convolutional methods is orders of magnitude faster than patch-based methods since evaluating the CNN output for a patch around every pixel in an image can be prohibitively slow. Patch-based methods, on the other hand, can draw large amounts of training data from small numbers of labeled images. However, neither approach takes into account the morphology of the objects to be segmented (e.g., contiguity) since the resulting probability map is thresholded in either case, leading to misclassifications in low probability areas. In cases where contiguous segmentation masks are needed to asses features such as branching patterns in retinal vasculature, the fully convolutional and patch-based methods can be suboptimal.



\section{Data and Methods}

\subsection{Data}

The DRIVE database consists of 40 images from a diabetic retinopathy screening program that have been divided into a training and test set, each containing 20 images. Human annotations for each image include a mask delineating the region of interest (RoI) and a manual segmentation of the vasculature (Figure \ref{data}, Left). We normalize each image to $[0,1]$ by dividing each pixel value by 255. We discard two images as outliers from the training set (labeled as ``23''and ``34'') since we found their corresponding manual segmentations to be noticeably different from the other 18 images in the training set. The remaining images in the training set are then randomly split into 15 training and 3 validation images. The results reported in this paper are averaged over 5 random splits. 

\subsection{Region Growing Convolutional Neural Network Architecture} \label{sec:network}

\begin{figure}[ht]
    \centering
    \vspace{-4mm}
    \includegraphics[width=\textwidth]{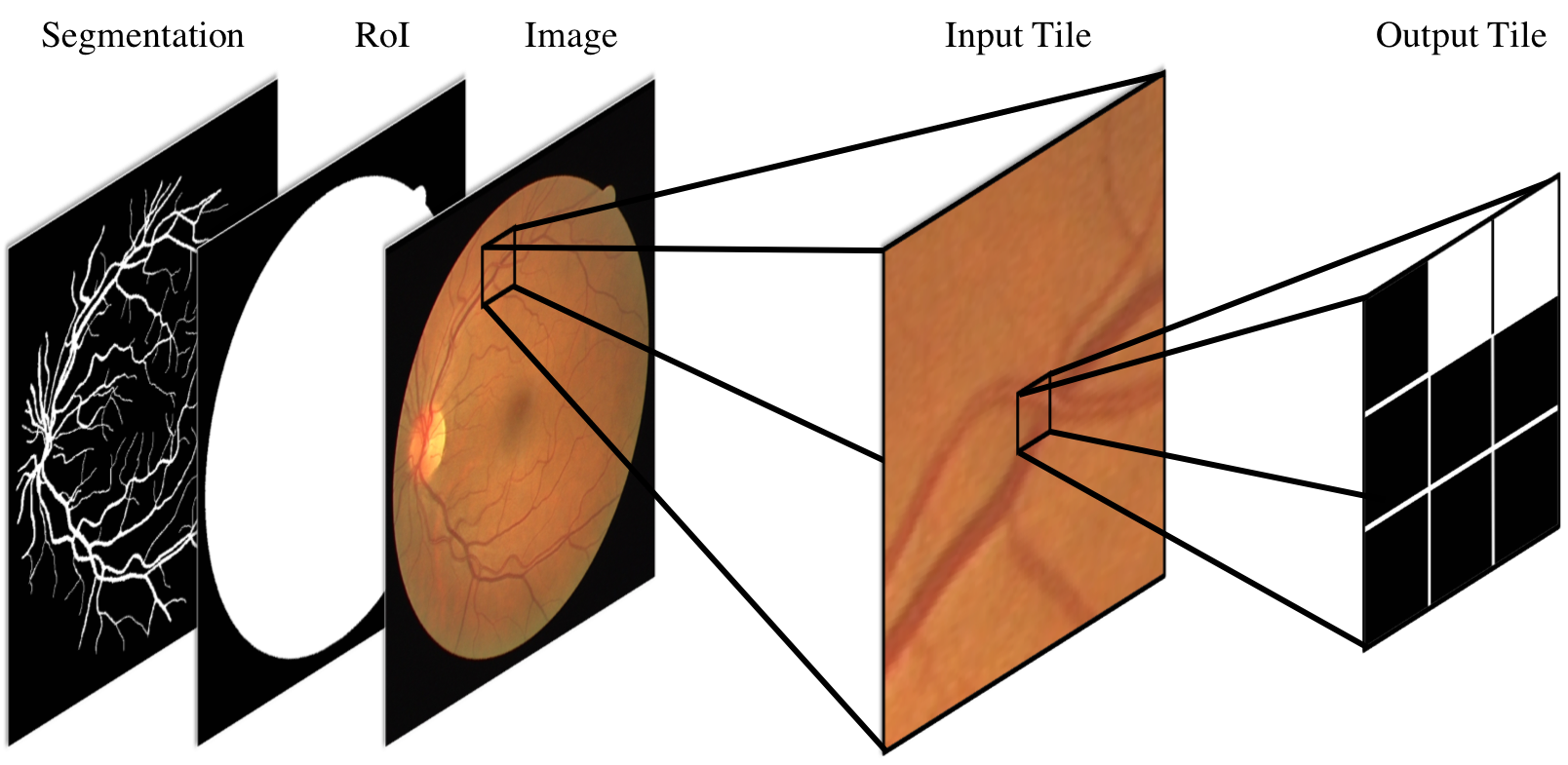}
    \caption{Left: Example segmentation and region of interest (RoI) for an image in the DRIVE database. Right: Example of an input and output for the convolutional neural network (RGCnet). Input: $80\times80$ image tile, Output: $3\times3$ tile of class labels corresponding to the center of the input tile in which 1=foreground (vessel) and 0=background.}
    \label{data}
    \vspace{-4mm}
\end{figure}

The input to the region growing CNN (RGCnet) is an $80\times80\times3$ tile of an image and the output is a $3\times3\times C$ tile of class probabilities where $C$ denotes the number of classes; $C=2$ for foreground (vessel) and background in these data (Figure \ref{data}, Right). In practice, the input tile size needs to be chosen based on the data set. We found that using an $80\times80$ tile for the DRIVE data provides RGCnet adequate context while maintaining computational efficiency.

RGCnet uses a 26-layer ResNet-style architecture with pre-activations \cite{he2016res,he2016preact} and 10\% dropout between residual blocks. The final convolution is followed by a fully connected layer that is reshaped into $3\times3\times2$ with a SoftMax activation along the channel axis. The network output is a prediction of the pixel classes in a $3\times3$ neighborhood in the center of the input tile. See Supplementary Figure 2 for the full network architecture. RGCnet is trained using a weighted pixel-wise cross-entropy loss and the Adam optimizer for 50 epochs with a batch size of 64. The weighting scheme is discussed in Section \ref{sec:IBE}. Input/output data were sampled from the training/validation splits to maintain balanced classes. For a given pixel in the RoI of an image, we count the number of foreground classes in its local $3\times3$ neighborhood, i.e., each pixel is assigned a count between 0 and 9. We then sample the images evenly for each count which yields roughly 500,000 training and 75,000 validation input/output tiles with a 50/50 split of foreground/background pixels. Training tiles are re-sampled every epoch while the validation sample is fixed. Input tiles are augmented using random 90$^\circ$ rotations and small brightness, saturation, contrast, and hue changes during training. The full-size images are zero-padded to ensure edge pixels could be used as centers of $80\times80$ input tiles. 

We note that with dropout, RGCnet could stagnate during the beginning of training. To remedy this, we pre-trained the network on a random sample of input/output training tiles, of which 50\% had center pixels in foreground and 50\% had center pixels in the background. After pre-training for one epoch, the training procedure then continued as described above. 

\subsection{Region Growing Algorithm} \label{sec:IBE}

\begin{figure}[ht]
    \centering
    \vspace{-4mm}
    \includegraphics[width=0.9\textwidth]{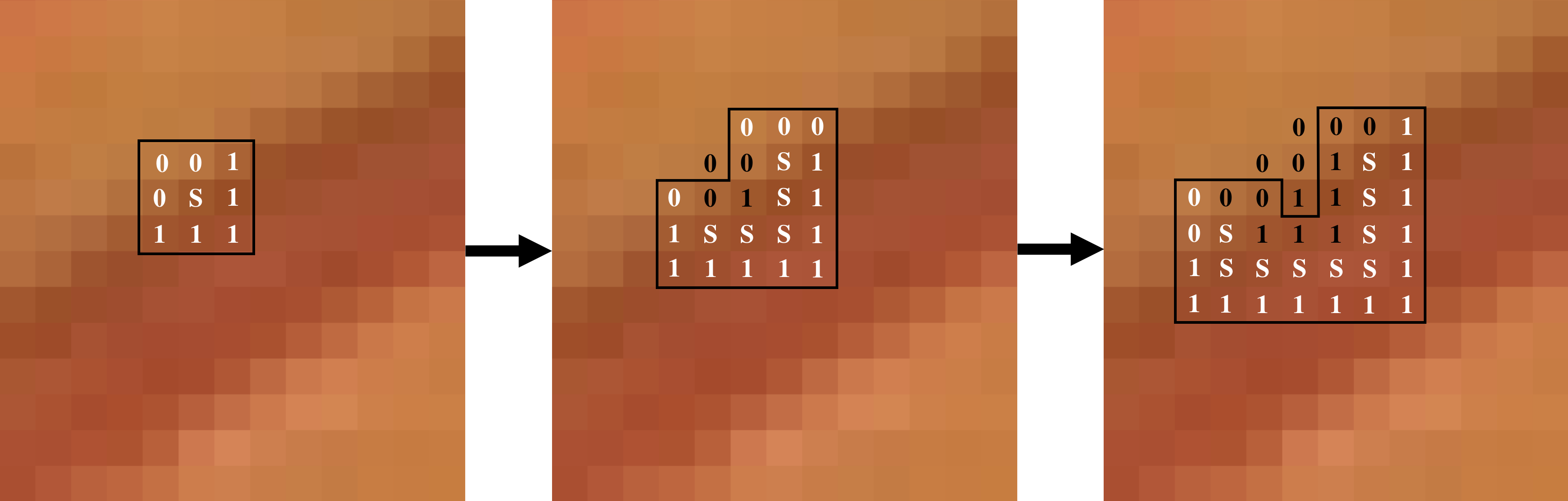}
    \caption{From left to right: Successive iterations of the region growing algorithm shown in a zoomed region of an image. A white ``S'' denotes the current pixel(s) in the sample set $\mathbb{S}$, a white ``0'' or ``1'' denotes the predicted pixel class in the current iteration (i.e., background and foreground, respectively), and a black ``0'' or ``1'' denotes the final prediction included in the mask. The black outline represents the area currently being processed in the iteration.}
    \label{iteration}
    \vspace{-4mm}
\end{figure}

Our segmentation algorithm uses RGCnet to iteratively spread predicted mask boundaries in each coordinate direction (Figure \ref{iteration}). The iteration begins by sampling a number of seed pixels at random in an image and adding them to a set $\mathbb{S}$. For images in the DRIVE database, seed pixels are sampled in the region of interest (RoI). For each pixel in $\mathbb{S}$, an $80\times80$ tile centered at the pixel location is input to RGCnet to predict class probability scores for the $3\times3$ neighborhood around the pixel; these computations are parallelized with a batch size of 100. Pixels in the RoI predicted as foreground are included in the segmentation mask and are added to the set $\mathbb{S}$ for the next iteration. Pixels that were input into RGCnet in the current iteration are removed from $\mathbb{S}$. We iterate until no new pixels are added to $\mathbb{S}$. This procedure ensures a contiguous segmentation from each seed since only pixels adjacent to previously classified foreground pixels can become part of the predicted mask. Note that pixels can receive more than one probability score in an iteration. This happens when multiple sample locations are adjacent because, in contrast to traditional patch-based methods, the output tile is larger than $1\times1$. To address this, we only include a pixel in the mask if its \textit{average} probability score exceeds some threshold, which we treat as a hyper-parameter. This hyper-parameter was chosen based on the optimal average evaluation metric score (Section \ref{evalmetric}) on images in the validation set. We optimized the threshold for each metric separately. We chose 10,000 random initial seed pixels from the RoI of each image to initialize the region growing algorithm for threshold optimization and test set prediction. On average, this corresponds to $>1$ seed for every $5\times5$ patch of the RoI. We found that 10,000 initial seeds was optimal for the DRIVE data in terms of accuracy and speed. See the Supplementary Movie for examples using different numbers of seeds.

Since the iteration relies heavily on the previous correct classification of foreground pixels (especially in low probability areas), it is crucial to distinguish pixels along the boundary between foreground/background. To focus RGCnet training on these regions, we multiplied the pixel-wise cross entropy loss by 5 for pixels along the contour of the ground truth segmentation mask and their adjacent background pixels.

\subsection{Baseline}
U-net \cite{ronneberger2015unet} was used as a baseline for comparison since it is widely adopted for semantic segmentation in biomedical imaging. We did not consider using instance segmentation methods, e.g., Mask-RCNN \cite{he2017maskrcnn}, as a baseline, since the DRIVE data represent a semantic segmentation task. We implemented U-net as described in \cite{ronneberger2015unet}, except that the data augmentations used were the same as those described above for RGCnet and we did not use a distance weighting scheme. We found that these augmentations yielded more accurate results for the DRIVE data than those reported in \cite{ronneberger2015unet}. The distance weighting scheme described in \cite{ronneberger2015unet} is not relevant to the DRIVE data semantic segmentation task, since there is primarily only one large object per image. U-net was trained using the same training/validation splits as RGCnet. Similar to RGCnet, the probability threshold for U-net was optimized for each evaluation metric over the validation set.

\subsection{Evaluation Metrics} \label{evalmetric}
We used three metrics to quantify segmentation accuracy: (1) Dice's similarity coefficient, (2) Jaccard index, and (3) mean symmetric surface distance (MSSD). The segmentation accuracy with respect to each evaluation metric was averaged over the 20 images in the test set. For segmentation masks $\Omega$ and $\Omega'$, the Dice coefficient, $2|\Omega\cap \Omega'|/(|\Omega|+|\Omega'|)$, and Jaccard index, $|\Omega\cap \Omega'|/|\Omega\cup \Omega'|$, are widely used accuracy metrics which strongly penalize false positives and false negatives but do not take into account the distance between the misclassified pixels and the ground truth. MSSD, however, may be a more biologically relevant metric as it can better capture the quality of the overall morphology of the predicted masks. MSSD is defined as
\begin{equation}
    \textrm{MSSD} = \frac{1}{|\Omega|+|\Omega'|}\left( \sum_{p\in \Omega}\textrm{d}(p,\Omega') + \sum_{p'\in \Omega'}\textrm{d}(p',\Omega) \right)
\end{equation}
where d$(p,\Omega')$ is the minimum Euclidean distance between a pixel $p\in \Omega$ and all pixels $p'\in \Omega'$ of the same class. Similarly for d$(p',\Omega)$. The distance between true positives is 0 and between false positives/negatives is $>0$. Thus, MSSD can represent the quality of the morphology since it penalizes misclassified pixels according to their distance from the ground truth mask. For example, a missing branch of the vasculature is penalized more heavily than over/under predicting the width of a vessel, whereas the Dice and Jaccard scores only account for the correct classification of each pixel. We note, however, that MSSD is only a good metric for morphology on the DRIVE data when the predicted mask is one large contiguous object. Otherwise, small artifacts scattered in the predicted mask can reduce the distances between pixels in the prediction and ground truth masks, weakening the morphological interpretability of the MSSD metric. For this reason, we also consider a post-processing step in which only the largest contiguous object in the predicted mask is used to calculate each metric.
\\
\textbf{Post-processing for the Largest Object: } 
Since our region growing algorithm was designed to better ensure contiguity, we tested whether only keeping the largest contiguous object in the predicted mask could increase segmentation accuracy.  However, since U-net does not ensure object contiguity, we found that keeping only the largest object in the predicted mask severely reduced U-net's segmentation accuracy with respect to all the evaluation metrics we considered here (Supplementary Figure 3). Therefore, we only present evaluation metrics for the case where all objects in the U-net mask are kept.
 
\section{Results} \label{sec:results}
\begin{figure}[!ht] 
    \centering
    \includegraphics[width=\textwidth]{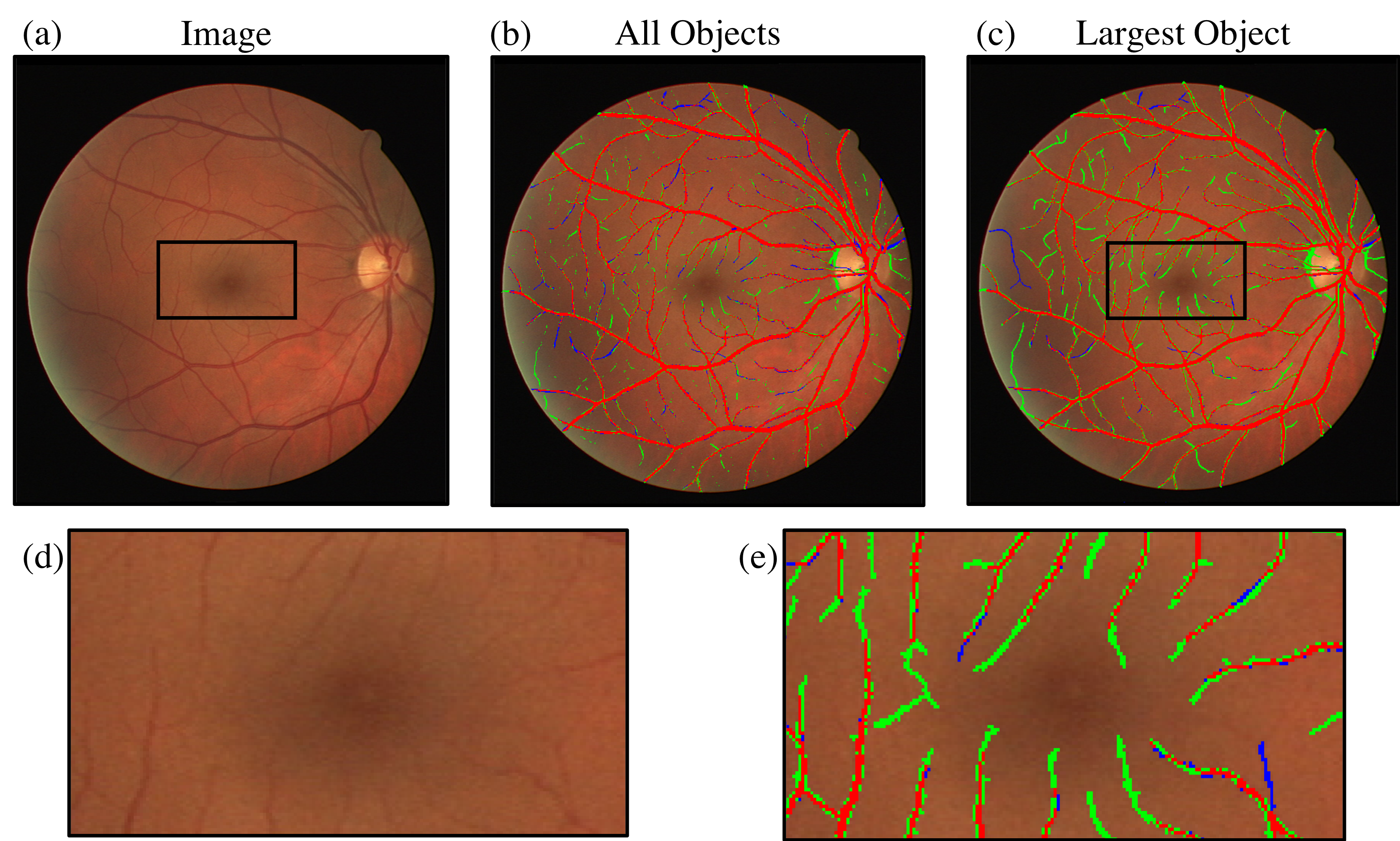}
    \caption{
        \textbf{(a):} An example test image in the DRIVE database. 
        \textbf{(b):} RGCnet segmentation mask with all objects kept. Predictions are labeled by color: {\color{red}Red}=True Positives, {\color{green} Green}=False Positives, {\color{blue} Blue}=False Negatives.  
        \textbf{(c):} RGCnet segmentation mask with only largest object kept.
        \textbf{(d):} Zoomed region from (a).
        \textbf{(e):} The corresponding zoomed region from (c). 
    }
    \label{tpr}
\end{figure}

We found that our region growing algorithm based on RGCnet was more accurate on average than U-net with respect to each evaluation metric, even when selecting only the largest contiguous object in the mask (Table \ref{tab:results}). The highest mean Dice and Jaccard scores were achieved by RGCnet when keeping only the largest object in the mask. This result is somewhat expected since we observed that keeping the largest object removed many artifacts in the predicted mask (Figures \ref{tpr}b and \ref{tpr}c). The opposite is true for MSSD. By keeping small artifacts in the mask, the mean symmetric distances between the prediction and ground truth are significantly reduced. However, the MSSD is a more biologically relevant metric when only one large object is in the predicted mask, since it can represent the quality of the overall predicted morphology. Thus, an average MSSD of $<1$ pixel when only keeping the largest object in the RGCnet mask strongly suggests that our region growing method accurately captures blood vessel morphology.
\begin{table}[ht]
    \centering
    \caption{Dice, Jaccard, and mean symmetric surface distance (MSSD) values on the testing set, averaged over 5 training/validation splits. Results are shown for the region growing algorithm (RGCnet) and the baseline method (U-net) with post-processing of keeping all objects (All) or only the largest object (Largest). \textbf{Bold} indicates best score.} 
    \begin{tabular}{c|c|c|c|c}
        Method & Post-processing & Dice & Jaccard & MSSD 
        \\ 
         &  & (mean $\pm$ std) & (mean $\pm$ std) & (mean $\pm$ std) \\ \hline
        RGCnet & All & 0.7936 $\pm$ 0.0022 & 0.6585 $\pm$ 0.0032 & \textbf{0.5143} $\pm$ 0.0142 
        \\
        RGCnet & Largest & \textbf{0.8042} $\pm$ 0.0007 & \textbf{0.6732} $\pm$ 0.0014 & 0.6976 $\pm$ 0.0389
        \\
        U-net & All & 0.7365 $\pm$ 0.0045 & 0.5836 $\pm$ 0.0057 &  1.1917 $\pm$ 0.1060 
    \end{tabular}
    \label{tab:results}
\end{table}

\section{Discussion}
We have shown that our CNN-based region growing algorithm is able to increase segmentation accuracy over U-net while preserving biologically realistic morphological features. Our method is faster than traditional patch-based methods, taking $\sim 30$ seconds per 584$\times$565 DRIVE database image on a NVIDIA GTX 1080Ti GPU. We observed that the region growing CNN can even detect vessels beyond the human annotated segmentation masks, revealing ambiguities/inaccuracies in the human-traced manual segmentations at finer scales. For example, see Figure \ref{tpr}d and false positives in Figure \ref{tpr}e which may actually be true positives. \\
\indent In future work, we will expand our approach to different neural network architectures, e.g., recurrent neural networks, and test its applicability across other biomedical imaging tasks. We will also consider extending this methodology to 3D segmentation tasks in which a 3D CNN will be used to determine the growth of regions.



%
%
%
\bibliographystyle{splncs04}
\bibliography{references}

\newpage

\section{Supplementary}

\begin{figure}[!ht] 
    \centering
    \includegraphics[width=\textwidth]{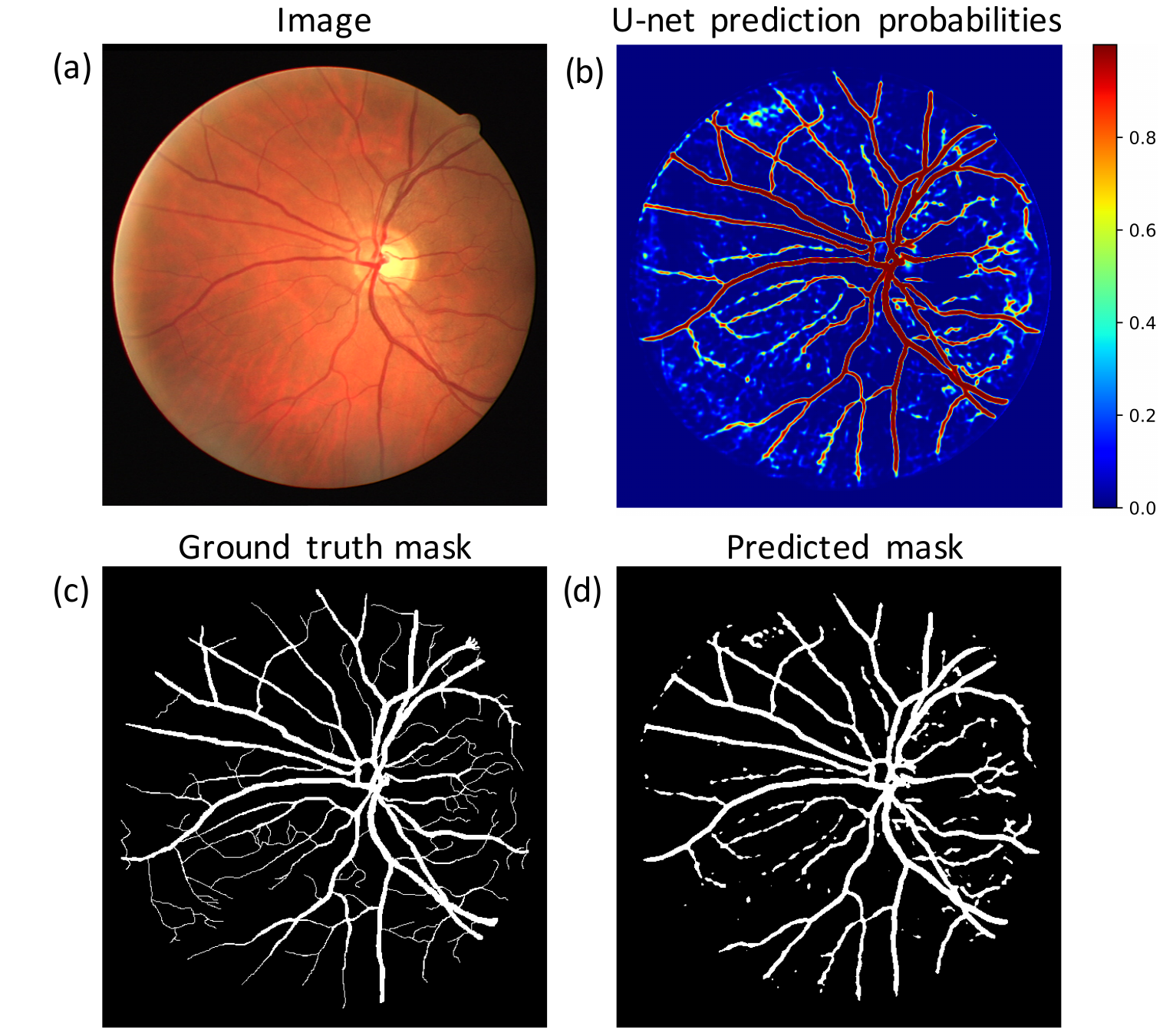}
    \caption*{\textbf{Supplementary Figure 1.} Lack of contiguity in the predicted mask from a fully convolutional neural network (U-net), resulting from patchy probability outputs.  \textbf{(a)}: Original image. \textbf{(b)}: Output probabilities from U-net. A probability equal to 1 corresponds to the foreground class (vessel), probability equal to 0 corresponds to the backgournd class. \textbf{(c)}: The ground truth mask for the image. \textbf{(d)}: The predicted mask resulting from thresholding the U-net probabilities.  See section 2.4 for details on U-net training and optimization of the probability threshold.  
    }
    \label{supp1}
\end{figure}

\begin{figure}[ht]
    \centering
    \includegraphics[width=\textwidth]{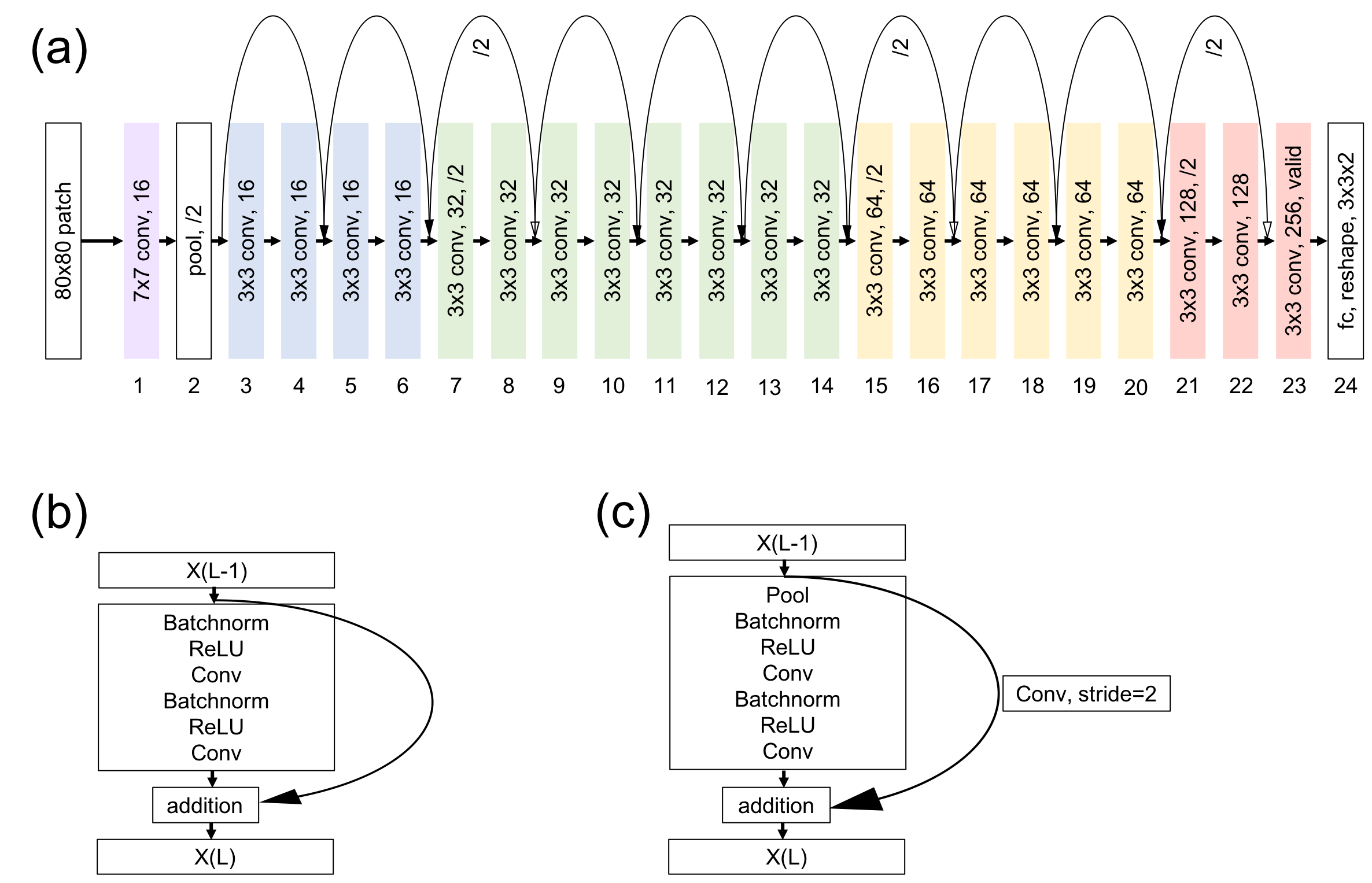}
    \caption*{\textbf{Supplementary Figure 2.} \textbf{(a)}: ResNet-style architecture with pre-activations. \textbf{(b)}: Example of a pre-activation residual module in which the input/output size does not change, \textit{e.g.}, 3-4, 5-6, 9-10, etc. in (a). \textbf{(c)}: Example of a pre-activation residual module in which the input/output size reduces by 2, \textit{e.g.}, 7-8, 15-16, 21-22 in (a).}
    \label{fig:resnet}
\end{figure}

\begin{figure}[!ht] 
    \centering
    \includegraphics[width=\textwidth]{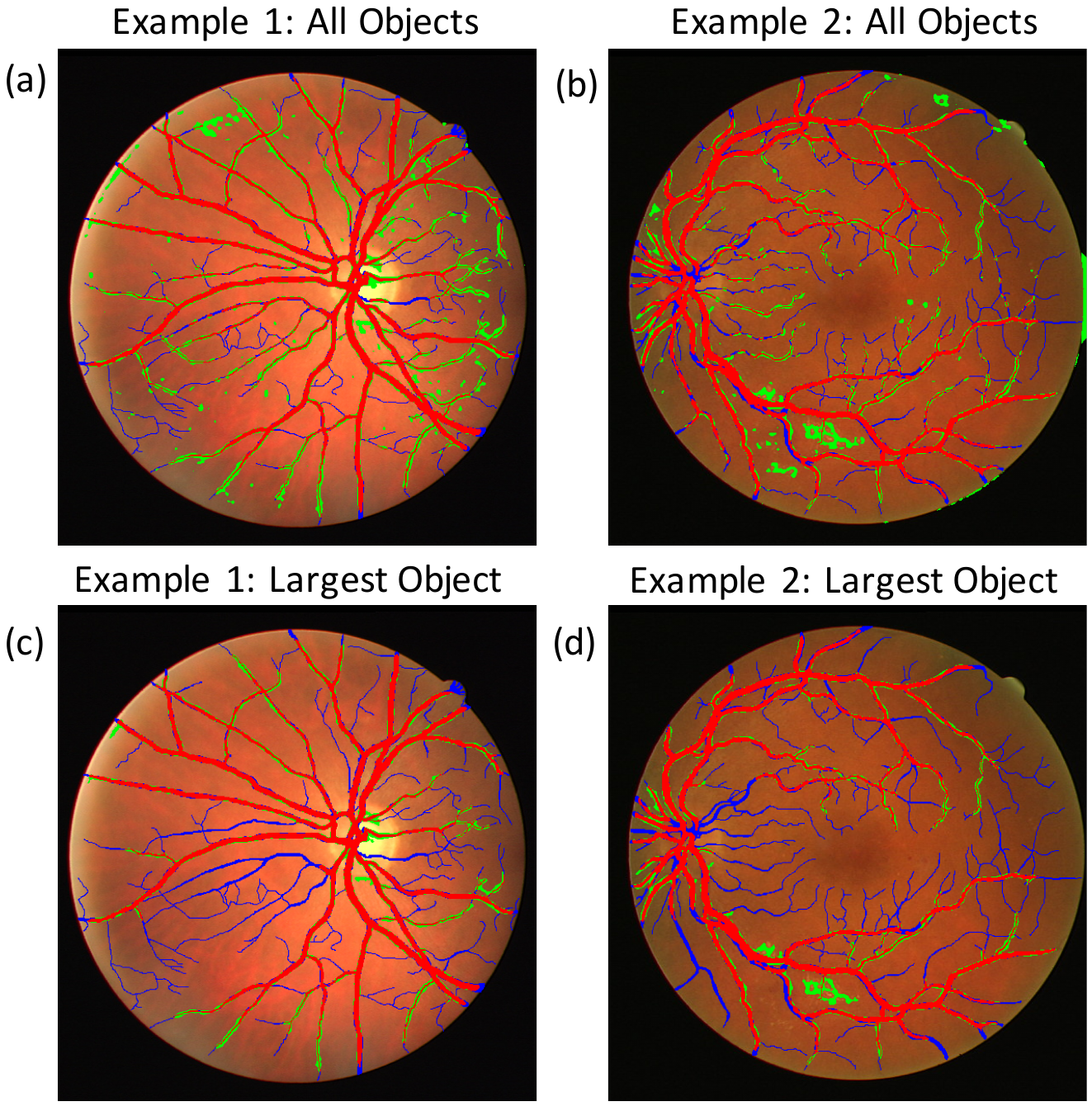}
    \caption*{\textbf{Supplementary Figure 3.} Keeping only the largest object did not improve U-net accuracy. U-net predictions for two representative example images from the test set are shown; Left: Example 1, Right: Example 2. \textbf{(a)} and \textbf{(b)}: Results from keeping all objects for the two examples. \textbf{(c)} and \textbf{(d)}: Results from keeping only the largest object for the two examples. Predictions are labeled by color: {\color{red}Red}=True Positives, {\color{green} Green}=False Positives, {\color{blue} Blue}=False Negatives. The mean symmetric surface distance (MSSD) values for these predictions are \textbf{(a)}=0.967, \textbf{(b)}=1.391, \textbf{(c)}=5.573, \textbf{(d)}=5.429. These example show that the MSSD is much lower when keeping only a single object. However, the MSSD is a less acccurate representation of morphological distance when keeping all objects, since many disconnected objects are needed to attain a low MSSD value.
    }
    \label{supp3}
\end{figure}

\end{document}